# Increasing the Network life Time by Simulated Annealing Algorithm in WSN with Point Coverage

Mostafa Azami[1], Manij Ranjbar[2], Ali Shokouhi rostami[3], Amir Jahani Amiri[4]

[1, 2] Computer Department, University Of Kurdistan, Sanandaj, Iran

m.azami@b-iust.ac.ir

[3,4] Electrical Department, University Of tamishan, Behshahr, Iran

a.shokouhi@b-iust.ac.ir

**Abstract**. *Since we are not able to replace the battery in a wireless sensor networks (WSNs), the issues of energy and lifetime are the most important parameters. In asymmetrical networks, different sensors with various abilities are used. Super nodes, with higher power and wider range of communication in comparison with common sensors, are used to cause connectivity and transmit data to base stations in these networks. It is crucial to select the parameters of fit function and monitoring sensors optimally in a point covering network. In this paper, we utilized an algorithm to select monitoring sensors. The selection is done by using a novel algorithm that used by simulated annealing. This selection takes remained energy into consideration. This method increases lifetime, decreases and balances energy consumption as confirmed by simulation results.*

**Keywords**: *Asymmetrical Sensor Networks, Energy, Lifetime.*

## 1 Introduction

THERE have been remarkable uses of cheap sensor networks with lower energy. The main challenge in designing wired and wireless systems emerges from the two significant sources: communication bandwidth, the energy of a system. Overcoming these restrictions requires designing new communication techniques to augment the needed bandwidth for each user and innovating energy-efficient protocols. Designs would be diverse in different applications due to the expectations of a system. For instance, the energy used in performance rounds must be optimized and the lifetime of a network ought to be maximized. Since battery replacement is not appropriate in many applications, low power consumption is a crucial requirement in these networks. The lifetime of a sensor can be efficiently increased by optimizing power consumption [3]. Power-efficient designs have found widespread uses in these networks. They are generally in the spotlight from the perspective of hardware, algorithm and protocol design. An approach to reduce energy consumption is to decrease the number of sensors in the sensing area in a way that target detection is guaranteed in the given area. If the network is scalable, the algorithm can optimally be employed to decrease the number of nodes [4].

Many concepts have been released for the word "Point" due to the variety in the number of sensors, their types and applications. Hence, point can be defined generally as a parameter of service quality in a wireless network. For example, it may be asked that what the quorum is for a wireless network to supervise a specific region or what the probability of detecting an incident is in a specific time interval. Besides, relations given for coverage find the weak points of a sense field and introduce amending designs to improve service quality of the network [2]. Some of the targets, with definite positions that must be controlled, are considered in the scenario of point coverage. Many of sensors are distributed randomly and they send achieved information to the central processor. Each target must be controlled by at least one sensor. One way of reducing energy consumption is decreasing the number of active sensors in a covered area by supposing that each sensor can control all existing targets in its own area. A method to increase the lifetime of a sensor network is dividing sensors into some separated sets. Each set must cover all targets. These separated sets are activated consecutively in a way that only one set is active at a moment. In addition to lifetime increase and decrease in the number of active sensors, the following provisions must be satisfied:

_the number of sensors must be in a way that target detection is guaranteed in the given area.

_each sensor must be able to connect to the center.

A sensor network can be deployed in three ways: random[3,16],controlled placement [2,6,7,10,14,18] and uniform distribution[19].

Random method is a method of network generation places a seed node at the origin . Sensors are usually randomly distributed over a wide sensing area .When the environment is unknown, random placement must be used and sensors may be randomly dropped from aircraft. sensors are initially position on distributed over the network domain that sensor locations are independent[3,16].

In [16] a method for achieving an appropriate coverage is suggested. It has a random distribution and sensor density is the variable parameter in the network. The optimum density of a sensor is achieved by defining an upper bound for the probability of point coverage and finding a relation between sensor density and the average of the area that is not covered.

In [3] a method for increasing network lifetime is proposed based on maximizing the number of sensor groups with random distribution and on the basis of the optimization of energy consumption. In this method, each node is allowed to be the member of more than one group which results in network lifetime increase.

Second type of sensor distribution is the controlled placement approach; sensors can be carefully deployed on a sensor field, if the terrain properties are predetermined. Consequently, the controlled placement can be planned to meet the requirements of various levels of services. For example, surveillance, target positioning and target tracking. If the planning process is subject to some resource constraints (such as, the deployment cost) and to achieve some specific goals, the sensor deployment will be considered as an optimization problem [2,6,7,10,14,18]. For example Carbunar and his colleagues have devised a method for decreasing energy consumption by determining the position of sensor nodes and decreasing their overlap area [2]. A method to decrease power consumption is released in [14] with controlled placement. At first, sensors are divided into two groups in this method. One of them is just active at a moment. Their activeness inactiveness approach is iterated periodically. The algorithm for decreasing the number of nodes can be employed optimally if the network is scalable. The energy is not distributed homogeneously in sensors for the networks with static distributions [6, 7]. In [10] an approach has been presented to access a scalable coverage which is utilized in the case of the existence of overhead and high computational complexity for boosting energy efficiency. A wireless network must be stable in the case of abnormality. In [18] a method is released for a three-dimensional coverage by considering a node to be able to be omitted or displaced. It's a self-amending method which optimizes the energy consumption of a network.

Third type of sensor distribution is the uniform distribution approach; in this approach Sensor nodes were randomly distributed using uniform distribution for their X and Y coordinates. In this topology, all sensor nodes were assumed stationary during the simulation. A uniform distribution is one for which the probability of occurrence is the same for all values of X. It is sometimes called a rectangular distribution. For example, if a fair die is thrown, the probability of obtaining any one of the six possible outcomes is 1/6. Since all outcomes are equally probable, the distribution is uniform. If a uniform distribution is divided into equally spaced intervals, there will be an equal number of members of the population in each interval. . In [19] a method to decrease the number of sensors and abate energy consumption is released on the basis of biological algorithms. Uniform distribution used in this method and our study is a privilege to other methods.

The limited energy supply of wireless sensor networks poses a great challenge for the deployment of wireless sensor nodes. In this paper, we focus on energy-efficient coverage with simulated annealing to increase lifetime. In scalable networks, optimal algorithms are used to reduce the number of nodes in the network. This paper, by using SA method has been discussed evaluation and optimal selection of sensor nodes to cover the environment, So that the sensors will have the maximum amount of silence. The advantage of this method compared to other methods is that selection sensors starting from an uniform state with high temperature and then with gradual cooling, we will discover Convergence to number of active minimum sensors and minimizing the energy. Thus reducing the energy consumption per lifetime will lead to maximize the network lifetime.

Both sensor deployment and energy conservation are key issues for WSNs. This work considers the problem of constructing an energy-efficient sensor network for surveillance and target positioning services using the uniform distribution approach. The design goals are to achieve target positioning as well as to prolong sensor network lifetime. To support positioning functionality, the sensor field must be completely covered and each unit in the field is discriminable. It requires deploying more sensors than to support surveillance functionality. However, to keep all sensors in active to support the target positioning service is not necessary and waste sensors' energy if intrusion events occur infrequently. Actually, the surveillance service is enough when there isn't any intruder in the sensor field.

The proposed grouping algorithm is presented in the second section. The third section is devoted to sensor network protocol design. Simulation results are discussed in the fourth section. The Last section is about total results.

## 2  Point coverage in WSN

The objective is to cover a set of points in the Point Coverage approach. Fig. 1 depicts a set of sensors which are deployed uniformly to cover a set of targets (the square nodes). The connected black nodes form a set of active sensors that is the result of the timing mechanism [14].

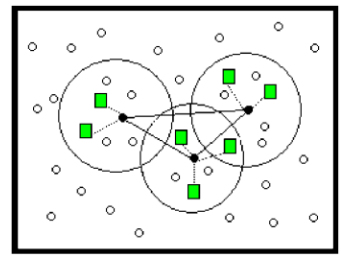

Fig. 1. Coverage Point

Timing protocol is one of the posed grouping protocols in sensor networks. This algorithm is based on the timing protocol of activity duration of the networks. It uses a two-phase mechanism (initiative and executive) and works on the basis of data communication in the shape of single-hop or multi-hop. Each group includes some super nodes, relay and monitoring sensors. In this protocol, group selection is done by using the fit function designed in the protocol.

In the initiative phase, some of nodes call themselves "sensor" and send their propagation messages to their neighbors. The second phase (executive) is known as the stable phase. In this phase, data reception or transmission is done from sensor nodes to relay nodes and from relay nodes to destination. Fig. 3 shows the schedule of the protocol operation.

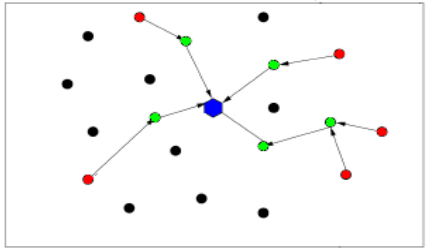

Figure. 2. Sample of selected set for covering targets - sensor nodes (red circles), relay nodes (green circles) and super node (blue hexagon)

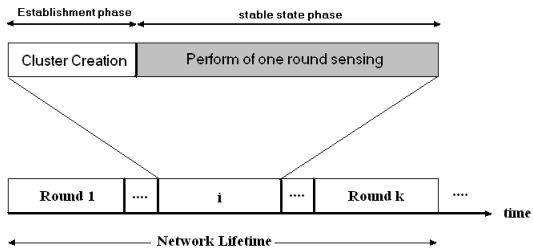

Figure. 3. Schedule of the protocol operation.

The nodes of a super node transmit data carefully, like LEACH algorithm, due to the schedule. The energy is saved by grouping in the rest time of inactive nodes. In grouping protocols due to the periodic circulation of active sensors, energy consumption is steady in the whole network. Hence, we used this feature in our paper. As shown in Fig. 2 each round includes two phases: initiative phase and executive phase. The initiative phase includes two parts. The former is devoted to the selection of monitoring sensors. The latter is for selecting relay sensors. It's so obvious that using super nodes increases the lifetime of the network.

## 2.1 Energy Model

The energy model for transmitting and receiving one bit of data has been assumed to be in accordance with LEACH energy model [14]. Assume that the distance between a transmitter and a receiver is d in the energy model mentioned above. If $d$ is more than $d_0$, the multi-path model (with path loss coefficient 4) is used. Otherwise the open space model (with pass loss coefficient 2) is used.

$$E_{Tx}(l,d) = E_{Tx-elec}(l) + E_{Tx-amp}(l,d) = \begin{cases} lE_{elec} + l\varepsilon_{fs}d^2 & d < d_0 \\ lE_{elec} + l\varepsilon_{fs}d^4 & d \geq d_0 \end{cases} \quad (1)$$

$E_{elect}$ is the required energy to activate the electrical circuit. $\varepsilon_{mp}$ and $\varepsilon_{fs}$ are the activation energies for power amplifiers in multi-path and open space cases, respectively and $q$ is represented as (2)

$$E_{Tx}(l,d) = p + qd^\alpha \quad (2)$$

On the receiver side, the consumed energy to receive one bit of data is as (3).

$$E_{Rx}(l) = E_{Rx-elec}(l) = lE_{elec} = p \quad (3)$$

In the presented asymmetrical networks, the initial energy of super nodes is assumed to be several times greater than the initial energy of typical sensors. The consumption energy of a monitoring node and a relay in each round are denoted by $Es1$ and $Ec1$, respectively.

## 3 Proposed Method

The problem is how to design a protocol to increase network lifetime and decrease energy consumption in the existence of these nodes. The benchmarks are trying to use the energy of common sensors to the most.

In covering networks, the physical positions of nodes and the number of the times of using them should be considered in protocol designing. How many times a sensor is used and also the distance between the selected node (in fact its relay path) and the super node has a crucial role for the energy consumption of that group. Therefore, we should be seeking for a relation between these two parameters and their energy consumption. At first, we state the problem and considered provisions. Then, simulation parameters that include timing algorithm based on the super nodes (for point coverage) will be explained. Our network has $N$ sensors named $S_1$ to $S_N$. We have M super nodes named $S_{u1}$ to $S_{uM}$ (M<N). The proposed timing algorithm divides the time interval into specific rounds and identical intervals $T_r$. Selected group is only active during the time $T_r$ and other groups are off during a round. The duration of a round $T_r$ can be computed by considering grouping time, the energy of groups, conjectured lifetime physical parameters and the types of common sensors used in the network.

## 3.1 Provisions dominating the network

The provisions of the network are listed below:
_ there are K targets with definite positions in the network composed of sensor nodes and super nodes. In the considered scenario, sensor nodes and super nodes are deployed uniformly. The schedule of the activities of sensor nodes must guarantee the following conditions after running the algorithm for network lifetime:

- There are targets $T_{a1}$ to $T_{ak}$ which must always be covered.
- There are nodes $S_1$ to $S_N$ which perform the monitoring task and are deployed uniformly.
- The super nodes $S_{u_1}$ to $S_{u_M}$ are deployed uniformly.

- There will be chosen sets of nodes $C_1$ to $C_j$. Each $C_j$ is a set of active nodes. It is constructed in each round by the protocol.
- Each set $C_j$ is necessary and sufficient to cover all the $k$ targets.

In fact, the objective is to divide sensor nodes into active and inactive groups. Active sensors must be able to do connectivity and coverage. The objective to use this algorithm is maximizing the groups to reduce energy consumption and increase network lifetime. In each performing round, it should be checked whether a node is active, a sensor node or a relay node.

- Each common sensor has the initial $E_i$ and a limited processing power. Unlike common sensors, super nodes have higher energy, greater lifetime and higher processing power.
- All super nodes are connected to each other so there is at least one path between two super nodes.
- Each active sensor exists in one of the $C_j$ groups and must be connected to a super node by relay nodes. It's connected to at least one super node through a path to transmit its own information to the super node.
- Sensor nodes possess initial energy $E_i$, communication range $R_c$, and sensing range $R_s$ ($R_c \geq R_s$).
- This selection must be local and distributed. Decision making is done by using the data of neighboring node with a fixed multi-hop distance.

*Definition 1:*
In defining point coverage, it should be said that when the Euclidean distance between our node and the target is less than or the same as $R_s$, the target is covered.

*Definition 2:*
Sensors can connect with each other or super nodes if their Euclidean distance is less than $R_c$.

*Definition 3:*
Network lifetime is defined as the time interval in which all $k$ targets are covered by a set of active sensor nodes that are connected to super nodes.

## 3.2 Sensing Nodes Selection Algorithm

As indicated before, designed grouping algorithm, executed in the beginning of each performance round, includes two sections. The first section is the selection of active nodes. The second section is attributed to gathering data from nodes and sending the data through relay nodes.

In the first section, one of $C_j$ groups is formed in a way that the above provisions are satisfied. When this group is active, all other nodes are inactive (Sleep Mode) and use little energy. They will be evaluated in the next phase. This evaluation is done by considering a series of the physical factors of sensors during a round.

### 3.2.1 System Specifications

The network is supposed to be in a squared environment. There are $T_{a_k}$ targets in the environment that must be covered to produce a connected covering network. $Tars_n$ Includes all the targets in the sensing domain of $S_n$. They are not covered by other apt nodes. The number of targets located in the sensing range of the node $S_1$ is shown by $m_1$.

The initial energy of common sensors is $E_i$ and the initial energy of super nodes is three times greater than $E_i$. The energy consumed in each round is called $Es_1$ and the consumed energy of a relay in each round is called $Ec_1$.

The first section including sensor node selection, fit function checking for evaluation and selecting active monitoring nodes takes $w$ time units (Second is the time unit here). The waiting time of the node $S_n$ is computed by a function measuring the physical parameters of the sensor $S_n$. Waiting time is stated as a coefficient multiplied by the whole time of a round by using the parameters of a node: remained energy, initial energy and the number of the targets observed in the range of a sensor. A sensor decides to sleep or remain awake after passing waiting time. If $E_n < E_{s_1} + E_{c_1}$ ($E_n$ is the remained energy of the sensor node $S_n$) then the node cannot be converted to a sensor node. So waiting time is not computed and $t_n$, the waiting time of the node n, is equivalent to w. It means that the node is not a sensor one. Otherwise, when $E_n > E_{s_1} + E_{c_1}$, $t_n$ is computed and inspected. When $t_n$ finishes and $Tars_n \neq \phi$, $S_n$ introduces itself as a sensor node and joins active nodes in the group. Then, new selected node says its position to the two-hop neighboring nodes. If there is a node such as $S_j$ at the end of the round that $Tars_n \neq \phi$ and $E_n < E_{s_1} + E_{c_1}$, the node sends the "no coverage" message to its super node. It means the lifetime of
the network is finished. At this time, a message containing "no complete coverage" is sent to super nodes and the network sends this message to the final monitoring destination.

### 3.2.2 Simulated Annealing Algorithm

SA algorithm is a random hill-climbing movement which shows sort of efficiency. To perceive it better, suppose that a ping-pong ball is inside the deepest hole of a rough surface. By shaking the surface, we can change the position of the ball to a minimum local place. But, it should not be much intense that make the ball far from a considered distance. SA's goal is to start with intense shakes (high temperature) and then reduce the intensity (Temperature Reduction).

Procedure Simulated Annealing:

```
S1 = Choose an initial solution
T = Choose an initial temperature
REPEAT
    REPEAT
        S2 = Generate a neighbor of the S1
    UNTIL s2 establish criteria
    ΔE = objective ( S2 ) – objective ( S1)
    IF ( ΔE > 0 ) THEN // s2 better than s1
        S1 = s2
    ELSE with probability EXP( ΔE/ T )
        S1 = s2
    END IF
    Decrease T
UNTIL meet the stop criteria
End
```

Instead of choosing the best movement, a random movement is chosen in this algorithm. If the movement improves the case, it will be accepted all the time. Otherwise, the algorithm accepts the movement with the probability value of 1 ($e^{\frac{\Delta E}{T}}$). Also, probability reduces due to temperature (T) reduction. Bad movements may be allowed when the temperature is high and decrease as the temperature reduces.

To reduce the temperature, T is multiplied by a coefficient between 0 and 1. Fast decrease of the temperature makes us encounter the problem of local optimality. So we choose a value near 1 for this parameter (for example 0.998).

- The steps of SA

    Start and Preparation:
    Entering problem information and adjusting parameters.
        1. Producing an answer near the current answer that accepted criteria.
        2. Evaluating this answer
            2-1) the neighbor is better than the current answer so go to a new answer.
            2-2) the probability is greater than the random number so go to a new answer. Otherwise get back to the first step.
        3. Updating the parameters of the problem and the algorithm. Move to the step 1.

Initially, the number of sensors is requested from the user. A binary chromosome generated randomly. In SA step, Optimized chromosome created from Primary chromosome. In this step, for create a neighbor chromosome, Two indexes will be chosen randomly. Then, these indexes, that represent two sensors, will be changed. Then Dikestra algorithm is used to determination relay nodes. If the sensors of a chromosome cover the whole network through this method reduce energy of active sensors and add the value 1 to the lifetime number, and this loop continues until $E_n < E_{s_1} + E_{c_1}$.

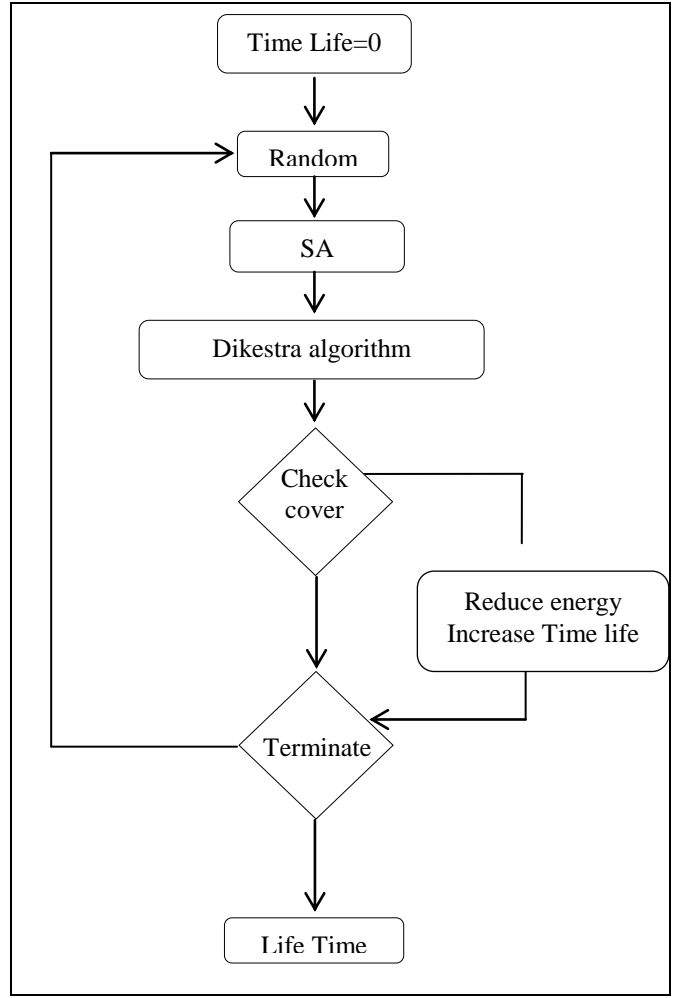

The values of simulation parameters are presented in table 1.

Table 1. Used values in simulation

| Parameter | Value |
| --- | --- |
| Network Size | 500 * 500 m |
| SNodes Location | Uniform Distribution |
| Nodes Location | Uniform Distribution |
| Nodes Initial Energy | 0.1 J |
| Super Node Initial Energy | 0.5 J |
| Communication Range | 90 m |
| Sensing Range | 60 m |
| Number of Nodes | 300 |
| Number of SNodes | 25 |
| Number of Target | 20 |
| Elect | 50 nJ/bit |

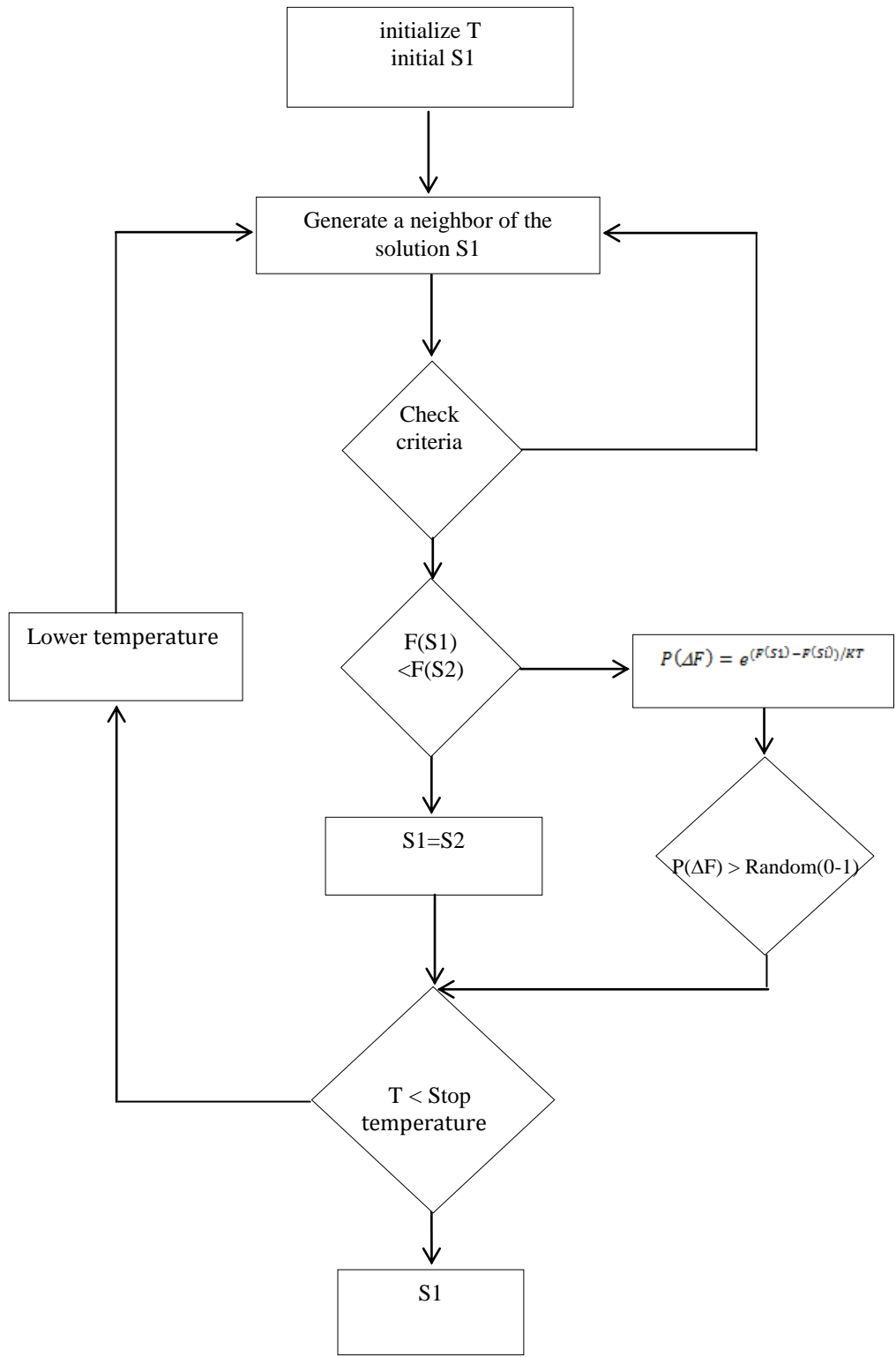

## 4. Results

In Figure 4 a sample of the response of the protocol to the selection of monitoring nodes, relay nodes and selected path is depicted for a round. The black squares are the targets. Typical /nodes and super nodes are shown by green-colored stars and red-colored circles, respectively.

To investigate the performance of the presented algorithm, C# has been used. In this section, the proposed algorithm is compared with the algorithms in [9, 1, 3, 11, and 12]. Compared its algorithms with other two algorithms. It was better than them.

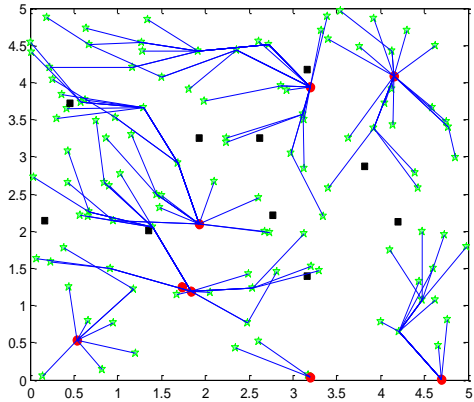

Figure 4. A sample of the response of the algorithm in a round

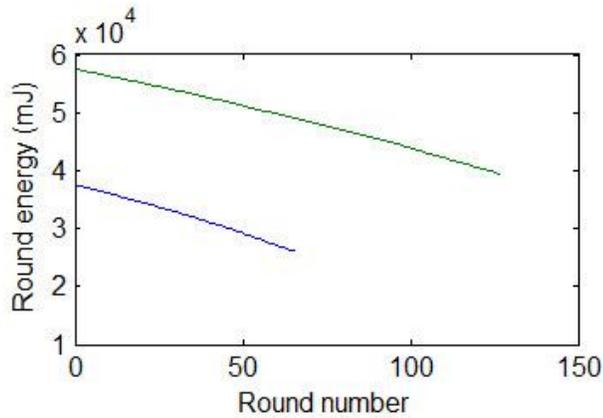

Figure. 5. Comparison of network lifetime for new algorithms

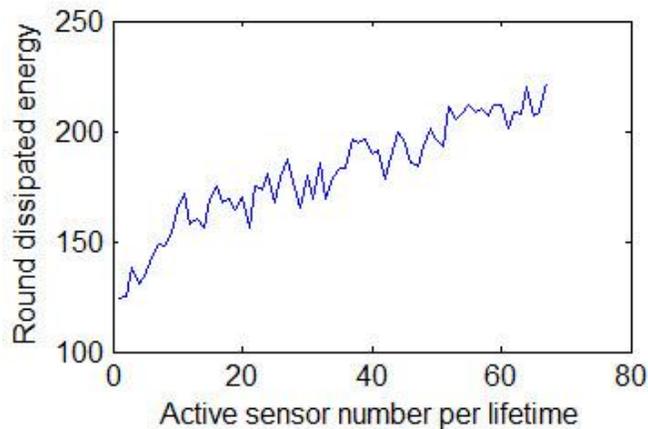

Figure. 6. Active sensor number per lifetime

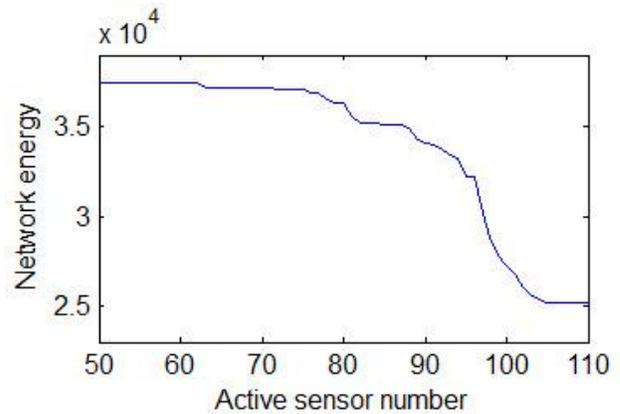

Figure. 7. Total Network Energy consumption with algorithms per active sensor number

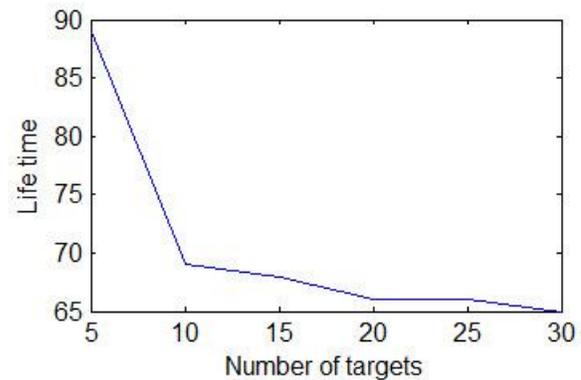

Figure. 8. Change of lifetime with increase of number of targets

As seen in the simulations, it shows that the SA algorithm is applied for selection of monitoring sensors in a point coverage network with increase in life time and reduction of average useful energy consumption. Amount of Increase in efficiency is significant than [11-13] Resources.

As it is shown in Table 1, this experiment was performed on a 500 x 500 environment; we distribute 300 nodes and 250 super nodes in this environment evenly to cover 20 targets which are distributed in the environment randomly. Communication range of each node is 90 m and its sensing range is 60 m also the amount of super nodes primary energy are 5 times greater than nodes. In this paper, to reduce the energy consumption and get smart sensor, two conditions have been considered:

- If the target was observed by several sensors, the sensor that sees more targets is selected for sensing.
- To reduce the activation interface nodes between target observer node and the nearest super node, Dijensra algorithm is used.

Simulations conducted in this paper for a point coverage network has been in C # language and has been tested by the MATLAB software. As shown in Figure 5, a simulation based on 300 sensors was done on the basis of the parameters in Table 1, which is reached to 67 rounds in lifetime of the network.

Also in Figure 5 we compare the change in network lifetime number for changes in nodes from 300 to 450 sensors; the green line is the result of 450 sensors and the blue line shows the state of 300 sensors.

Calculations done are based on implementation of the algorithm in 10 times and averaging the values of the lifetime of this number. The test has reached to 130 rounds for 450 sensors and similar condition with Table 1. According to Figure 6, the chart of average energy consumption, the consumed energy per round increases reasonably and amount of consumption has gentle ascending trend.

Figure 7 shows the number of active sensors for complete coverage network. Amount of Increase in active sensors has been about 76 sensors in the beginning of network to 108 sensors at end of network lifetime.

Before activation of 76 sensors, the amount of network remaining energy is fixed and by the first lifetime, network remaining energy is decreased. As it is evident in the figure, Energy reduction development has been slow in the early implementation of the program, but gradually according to energy completion of some sensors we had to use more sensors to cover the entire environment that leads to drastic reduction of network remaining energy in the last lifetimes.;

The lifetime of network has drawn In Figure 8 by changing the number of objectives and implementation of algorithms to 10 times, that comparing with [11-13] it is seen that the SA for a point coverage network has identical condition and in some cases also has the advantages to 10 percent. According to increase in lifetime factor, the increase has also seen to 14 percent compared to other researches.

## 5. Conclusions

In this paper, we presented a method to select active sensors in each round in asymmetrical wireless covering networks. In previous methods, this selection was dependent on the parameters of each sensor. In the proposed algorithm, this selection is according to the contest between neighboring nodes. Energy consumption in this method is more balanced than other similar methods. Generally, diagrams illustrate that the proposed method for the selection of monitoring sensors in point coverage is more energy-efficient than GSA, EEDG and EDTC. Also, it causes a longer lifetime.

According to simulation results, it is observed that this algorithm is well acted to solve this problem and optimization of a wireless sensor network in large-scale and it is able to provide a good and implementable response for network design and we can achieve better energy efficiency by organizing the network nodes and classifying them. Higher performance of network leads to increasing network life time.

Using SA algorithm, the categories were chosen so that the energy consumption in the network is minimized. Creating balance and uniformity in energy consumption of nodes and prolonging network life time is the Outcome of using the algorithm.

## References

[1] Awada. W and Cardei. M,"Energy Efficient data Gartering in heterogeneous Wireless Sensor networks, IEEE, WiMob,2006.

[2] Carbunar. B, Grama. A and Vitek. J, "Distributed and Dynamic Voronoi Overlays for Coverage Detection and Distributed Hash Tables in Ad-hoc Networks," (ICPADS 2004), Newport Beach, CA, pp. 549-559, July 2004.

[3] Cardei. M, Thai. M. T, Wu. Y. Li , W,"Energy-Efficient Target Coverage in Wireless Sensor Networks,",IEEE INFOCOM, 2005.

[4] Carle. J, Simplot. D," Energy efficient area monitoring by sensor networks," IEEE Computer 37 (2) pp 40–46, 2004.

[5] Chen. C & Ma. J, "Designing Energy-Efficient Wireless Sensor Networks with. Mobile Sinks," ACM International Workshop on (WSNA), pp. 343–349, 2006.

[6] Chen. C, Yu. J, "Designing Energy-Efficient Wireless Sensor Networks with Mobile Sinks," , ACM, USA. 2006.

[7] Huang. C.-F and Tseng. Y.-C, "The coverage problem in a wireless sensor network," in Proc. ACM International Workshop on (WSNA), pp. 115–121, 2003.

[8] Kim. J, Kim. D, " Energy-efficient Data Gathering Techniques Using Multiple Paths for Providing Resilience to Node Failures in Wireless Sensor Networks,", JOURNAL OF COMMUNICATIONS, VOL. 1, NO. 3, JUNE 2006.

[9] Liu. Zh, "Maximizing Network Lifetime for Target Coverage Problem in Heterogeneous Wireless Sensor Networks,", MSN 2007, LNCS 4864, pp. 457–468, 2007.

[10] Lu.J, Wang.J, and Suda.T, "Scalable Coverage Maintenance for Dense Wireless Sensor Networks," EURASIP Journal on Wireless Communications and Networking Volume, 2007.

[11] Rostami, A. S., Bernety. H.M, Hosseinabadi. A. R, "A Novel and Optimized Algorithm to Select Monitoring Sensors by GSA", ICCIA 2011.

[12] Rostami, A.S.; Tanhatalab, M.R.; Bernety, H.M.; Naghibi, S.E. Decreasing the Energy Consumption by a New Algorithm in Choosing the Best Sensor Node in Wireless Sensor Network with Point Coverage, (CICN), 2010.

[13] Rostami, A.S.; Nosrati, K.; Bernety, H.M, Optimizing the parameters of active sensor selection and using GA to decrease energy consumption in point coverage wireless sensor networks, (WCSN), 2010.

[14] Slijepcevic. S and Potkonjak. M., "Power efficient organization of wireless sensor networks," Proc. IEEE International Conference on Communications, pp. 472–476, 2001.

[15] Tian. D and Georganas. N. D, "A node scheduling scheme for energy conservation in large wireless sensor networks," Wireless Comm. Mob. Comput., vol. 3, pp. 271–290, 2003.

[16] Wang. B, Chua. K. Ch, Srinivasan. V, and Wang. W, "Information Coverage in Randomly Deployed Wireless Sensor Networks," IEEE Transactions Wireless Sensor Networks, Vol. 6, No. 8, Aug, 2007.

[17] Wang. W, inivasan. V. S, K.Ch. Chua, B. Wang, "Energy-efficient Coverage for Target Detection in Wireless Sensor Networks,", 2007 ACM.

[18] Watfa. M. K, Commuri. S, "An energy efficient and self-healing 3-dimensional sensor cover," (IJAHUC), Vol. 3, No. 1, 2008.

[19] Xu.Y and Yao.X, "A GA Approach to the Optimal Placement of Sensors in Wireless Sensor Networks with Obstacles and Preferences," IEEE CCNC, 2006.